\documentclass{elsart}
\usepackage{natbib}
\def \sax {BeppoSAX}

\def \nh {N${\rm _H}$}

\def \hcm {\hbox {\ifmmode $ atom cm$^{-2}\else atom cm$^{-2}$\fi}}

\def \chisq {$\chi ^{2}$}
\def \rchisq {$\chi_{\nu} ^{2}$}
\def\approxgt{\mathrel{\hbox{\rlap{\lower.55ex \hbox {$\sim$}}
        \kern-.3em \raise.4ex \hbox{$>$}}}}
\def\approxlt{\mathrel{\hbox{\rlap{\lower.55ex \hbox {$\sim$}}
        \kern-.3em \raise.4ex \hbox{$<$}}}}

\begin{document}
\runauthor{A. Orr}
\begin{frontmatter}

\title{
NLS1s and Sy1s: A comparison of ionized X-ray absorber properties 
}    

\author{Astrid Orr\thanksref{FN}}

\address{ESA/ESTEC, Keplerlaan 1, Postbox 299, NL- 2200 AG Noordwijk}
\address{IAAT, 64 Waldh\"auserstr., D-72076 T\"ubingen}

\thanks[FN]{A.O. acknowledges a fellowship of the Swiss
National Foundation for Scientific Research}

\begin{abstract}
The first results 
from a systematic study of warm absorbers 
in NLS1s  using \sax\ public archive data are presented here. 
We confirm ASCA results showing that a
warm absorber, as modeled by two oxygen K-shell absorption edges, is less
frequent in NLS1s than in broad line (BL) Sy1s ($\sim$20\% 
versus $\sim$50\%).
However, our study suggests that the ionization state of NLS1s
is {\em not} lower than that of BLS1s, as opposed to  the ASCA-based 
results.

The soft excess temperatures of our sample, 
when fitted with  blackbody emission models, lie within a small range of values
($\sim$0.02--0.15 keV in the rest frame) 
with no marked dependence on source luminosity. This is in agreement 
with ASCA-based findings for NLS1s and early results from IUE-ROSAT
BL Sy1 observations.

\end{abstract}

\begin{keyword}
AGN; NLS1; X-rays;
absorption
\end{keyword}

\end{frontmatter}


{\bf The \sax\ NLS1 sample}\\
 
The spectra of our sample were obtained with
the LECS \cite{p97}, MECS \cite{b97}, HPGSPC \cite{m97} and PDS
\cite{f97} instruments on-board \sax. 
Because of its good low-energy resolution and effective area down  
to 0.1 keV, \sax\ is very efficient for the study of the complex  
spectral features in the soft X-rays. The broad-band sensitivity of \sax\ 
also allows a much more accurate measurement of the underlying continuum than 
is possible with other X-ray observatories. 
Our  sample is composed of all the publicly available 
\sax\ NLS1 spectra before December 1999, plus the proprietary data for one 
source (Mrk~335).
Table \ref{source_para} lists  the sources with their X-ray luminosities
and redshifts. The X-ray luminosities span more than 4 orders of magnitude,
from the narrow-line QSO PKS~0558$-$504  down 
to NGC~4051 in its ultra-low state.
The data were processed using the SAXDAS 2.0.0 data analysis package
and the spectra were rebinned in order to allow the use of the $\chi^2$ statistic.
Where available, data were selected in the energy ranges
0.1--4.0~keV (LECS), 1.8--10~keV (MECS), 8.0--34~keV (HPGSPC),
and 15--100~keV (PDS). 
Because the spectra of IRAS 13224$-$3809 and the low state of NGC~4051  have 
a relatively low  signal to noise ratio,  
only data below 10 keV were considered for these two objects.

\begin{table}
\caption{The exposure times are for the MECS.
A value of H$_0=50$ km s$^{-1}$ Mpc$^{-1}$ is assumed. The
units  for \nh,  t$_{\rm exp}$ and L$_{0.1-10}$ are $10^{20} {\rm cm}^{-2}$, 
$ 10^4{\rm s}$ and  erg s$^{-1}$ respectively }
\label{source_para}
\begin {center}
\scriptsize
\begin{tabular}{l l l r l l}
\hline
Source Name & redshift & \nh &  date & t$_{\rm exp}$  & 
L$_{0.1-10}$ \\ \hline
NGC 4051-low    & 0.00242&  1.31 & 9-11 May 98  & 2.6 & $0.87 \times 10^{41}$ \\ 
NGC 4051-high   &        &       & 28-29 Jun 98 & 1.2 & $1.75 \times 10^{42}$ \\ 
RE~J~1034+39    & 0.04244&  1.43 & 18-19 Apr 97 & 4.3 & $3.8  \times 10^{43}$ \\ 
Mrk 766         & 0.013  &  1.77 & 17-18 May 97 & 7.8 & $4.9  \times 10^{43}$ \\ 
Mrk 335         & 0.025  &  3.77 & 10-12 Dec 98 & 8.9 & $8.0    \times 10^{43}$ \\ 
IRAS 13224$-$3809 & 0.0667 & 4.79  & 29-30 Jan 98 & 2.2 & $1.59 \times 10^{44}$ \\ 
Ark 564         & 0.024  & 6.4   & 14 Nov 97    & 2.4 & $2.8  \times 10^{44}$ \\ 
TON~S~180       & 0.062  & 1.5   & 3 Dec 96     & 2.3 & $3.1  \times 10^{44}$ \\ 
PKS 0558$-$504    & 0.137  & 4.39  & 18-19 Oct 98 & 6.4 & $4.4  \times 10^{45}$ \\ 
\hline
\end{tabular}
\normalsize
\end{center}
\end{table}

{\bf Fit results for the warm absorber}

All our fit models include Galactic absorption as well as 
factors to allow for flux normalization 
uncertainties between the instruments. 
These factors were constrained  within 
their standard ranges \cite{f:99}.   
Uncertainties are quoted as 90\% confidence intervals for one
parameter of interest.

The single power-law model does not give a good fit to any of the spectra
except PKS~0558$-$504, where the fit is marginally acceptable 
(\chisq = 1.19, 109 dof).
A soft excess, in the form of either power-law or blackbody emission,
gives significant improvement to the fit statistics of {\em all} spectra. 
In 4 spectra (IRAS~13224-3809, Mrk~335, Mrk~766, NGC~4051-high),  
the soft excess is slightly better fit by a blackbody component than by a power law.
In the remaining 5 spectra, the opposite is true.
We then
tried adding a Fe K$\alpha$ line  
and  2 K-shell absorption edges (O{\sc vii} and 
O{\sc viii}) to the power-law plus blackbody continuum.
The edge energies are fixed at ${\rm E} = 0.74$
and 0.87 keV in the rest frame and the line  at E$_{\rm rest}=6.4$ keV with 
$\sigma = 0.1$ keV. The fit results are listed in Table \ref{fit_results}. 
Only two spectra of our sample  (Mrk~335 and Mrk~766) 
show significantly improved fits (confidence level $>$ 95 \%) 
with the addition of these three spectral features. 
Three others show a marginal improvement (TON~S~180, NGC~4051-high and Ark~564).
The remaining four spectra show no improvement in the fit statistics. 
In  all 9 spectra, we obtain only upper limits for $\tau_{\rm O7}$. 
Likewise, in 7 spectra there are upper limits for $\tau_{\rm O8}$.
Only two spectra (Mrk~335 and Mrk~766) have definite, albeit low, values for
$\tau_{\rm O8}$.

\begin{table}
\caption{Two edge plus Fe K$\alpha$ line fit results. 
NA = not available}
\label{fit_results}
\begin {center}
\scriptsize
\begin{tabular}{l l l l l}
\hline
Source Name &  $\tau_{\rm O7}$ & $\tau_{\rm O8}$  & EW(Fe K$\alpha$) eV & \rchisq 
(dof) \\  \hline
NGC 4051-low    &  $< 0.34 $ &    NA     & NA  & 1.44 (30)\\
NGC 4051-high   &  $< 1.68 $ &  $< 1.65$ & $208 \pm 135$ & 1.12 (147) \\
RE~J~1034+39    &  $< 0.08 $ &  $< 0.10$ & $< 624$ & 1.22 (124)\\
Mrk 766         &  $< 1.53 $ &  $0.20 \pm 0.16$ & $102\pm_{42}^{60}$ & 1.34 (158)\\
Mrk 335         &  $< 0.07 $ &  $0.24 \pm_{0.21}^{0.22}$ & $228 \pm 82$ & 1.43 (138) \\
IRAS 13224$-$3809 &  $< 1.08 $ &  $< 0.49$ & $<813$&  1.48 (11)\\
Ark 564         &  $< 0.21 $ &  $< 0.08$ & $299\pm 122$ & 1.20 (156)\\
TON~S~180       &  $< 0.12 $ &  $< 0.11$ & $306\pm 214$ & 1.30 (187)\\
PKS 0558$-$504    &  $< 0.09 $ &  $< 0.06$ &$<104$ & 1.03 (104)\\
\hline
\end{tabular}
\normalsize
\end{center}
\end{table}

We have also tried fitting a warm absorber model ({\tt absori} in {\sc Xspec})
instead of the two O edges.
In this case, 5 spectra out of 9 show a significant improvement in quality 
of fit over the 2-edge fits: Mrk~335, TON~S~180, RE~J~1034+39, Ark~563 and 
NGC~4051-low. For the first three in this list, the improvement in \chisq\
has an F-statistic $>15$. Among the remaining spectra, 
it is evident that PKS 0558$-$504 has no warm absorber, and that 
{\tt absori} brings no significant improvement over the 2-edge fit 
for NGC~4051-low or Mrk~766. However, Mrk~766 most likely does have 
two O absorption edges. 
In the cases of NGC~4051-low and IRAS 13224$-$3809,
the poor quality of the spectra does not allow one to conclude 
anything about the
presence of soft X-ray features.

{\bf Comparisons with ASCA NLS1 results and with BLS1s}

Recent studies based on ASCA data have shown that warm absorbers are 
a common spectral component of BL type 1 AGNs \cite{re:97} \cite{ge:98}.  
Reynolds \cite{re:97} detected O{\sc vii} and O{\sc viii}
K-shell absorption edges in 50\% of his sample of 24, with $>$37\%
having optical depths greater than 0.2 for one or both edges. 
The average optical depths for O{\sc vii} and O{\sc viii}
in the Reynolds sample are $0.29\pm0.07$ and $0.18\pm 0.06$, respectively
\cite{le:99}.
These type 1 AGN results are in sharp contrast with the  
values from our \sax\ NLS1 sample (see Table \ref{fit_results}), 
that has only 2 clear detections of O absorption edges
in a total of 9 spectra, corresponding to $\sim$20\%.
A similar low proportion of $\sim$26\% was found by Leighly \cite{le:99}
in her study of NLS1 ASCA spectra, where 6 spectra from 5 sources 
(out of 23) have significant O absorption edge detections. 
We note that all the
sources of our sample are included in the larger ASCA sample
\cite{le:99}, but only one
of our two sources with significant edge detections (Mrk~766) is among
Leighly's five ``warm-absorbed'' NLS1.

The average optical depths for O{\sc vii} and O{\sc viii}
in the ASCA sample \cite{le:99} are $0.19\pm 0.14$ and $0.053\pm0.020$, 
respectively\footnote{The soft excess is modeled by a blackbody, as with our
\sax\ sample.}. 
The low value of $\tau_{\rm O8}$ suggests that the 
ionization state of NLS1s is  lower than that of BLS1s \cite{le:99}. 
However, looking at Table \ref{fit_results}
we see that {\em the \sax\ results do not support this conclusion}. 
The ASCA sample includes
objects which may have dusty warm absorbers 
(e.g. IRAS~13349+2438, IRAS~17020+4454 and IRAS~20181$-$2244). 
If the effects of the dust are not taken in to account correctly
one can derive warm absorber ionization states which 
are too low \cite{ko:98}.
Furthermore, it is a common assumption that the central black holes of 
NLS1s accrete at very high rates. In this case, a high power output 
is expected that could cause the ionization state of the warm absorber 
to be correspondingly high.

ASCA-based studies of NLS1s \cite{le:99} \cite{v:99} 
have shown that certain sources (e.g Ark~564) 
have broad absorption features in the  1.1-1.4 keV range.
We find that we can reproduce the \sax\ results on Ark~564 published
by  Comastri et al. \cite{co:98}, 
but not those of  Vaughan et al. \cite{v:99} based on ASCA data. 
In other words, a weak emission line or absorption edge may be present, but 
probably no absorption line as in \cite{v:99}.
And when a warm absorber ({\tt absori}) is included, 
the need for such a line is no longer very strong.
We have not yet tested the other sources in our sample
for such features, but it is clear that 
no large residuals are present between 1.1-1.4 keV
in our {\tt absori} fits.

The EWs of the Fe K$\alpha$ line in our sample are low and, within the
uncertainties, compatible with values expected in typical Sy 1 BLR 
conditions, i.e. column densities of 10$^{22-23}$ cm$^{-2}$, 
leading to EWs of $\sim$50--150 keV.
The soft excess temperatures of our sample, 
when fitted by  blackbody emission, lie within a small range of values
($\sim$0.02--0.15 keV in the rest frame) 
with no marked dependency on source luminosity. This is in agreement 
with ASCA-based findings for NLS1s \cite{le:99} \cite{v:99} and 
also IUE-ROSAT BL Sy1 observations \cite{wa:94}.

\end{document}